\input  phyzzx
\input epsf
\overfullrule=0pt
\hsize=6.5truein
\vsize=9.0truein
\voffset=-0.1truein
\hoffset=-0.1truein

%
% *** variable definitions ***
%

\def\IC{{\ \hbox{{\rm I}\kern-.6em\hbox{\bf C}}}}
\def\IR{{\hbox{{\rm I}\kern-.2em\hbox{\rm R}}}}
\def\IZ{{\hbox{{\rm Z}\kern-.4em\hbox{\rm Z}}}}

\def\sIR{{\hbox{{\sevenrm I}\kern-.2em\hbox{\sevenrm R}}}}

\def\mm{matrix model}
\def\dob{D0 branes }

%
% *** hyphenations ***
%
\hyphenation{Min-kow-ski}

\rightline{SU-ITP-96-12}
\rightline{April  1996}
\rightline{hep-th/9611164}

\vfill

%
% *** title ***
%
\title{T Duality in M(atrix) Theory and S Duality in field theory}

\vfill

%
% *** author list ***
%

\author{Leonard Susskind\foot{susskind@dormouse.stanford.edu}}

\vfill

\address{Department of Physics \break Stanford University,
 Stanford, CA 94305-4060}

\vfill

%
% *** abstract ***
%

The  \mm \ formulation of M theory can be generalized to compact
 transverse backgrounds such as tori. If the number of compact directions
is K then
the matrix model must be generalized to K+1 dimensional super Yang Mills
theory on a
compact space.  If K is greater  than or equal to 3, there are  T  
dualities
which
which require highly nontrivial identifications between different  SYM
theories. In
the simplest case we will see that the requirement reduces to the well
known electric-
magnetic duality of N=4  SYM theory in 3+1 dimensions.

% *** PACS categories ***
% PACS categories: 04.70.Dy, 04.60.Ds, 11.25.Mj, 97.60.Lf
\vfill\endpage

%
% *** References
%

%Banks et al, "M Theory as a Matrix Model: a Conjecture
\REF\bfss{T.~Banks, W.~Fischler, S.~Shenker and L.~Susskind,  
hep-th/9610043}
% Susskind, "Some Speculations About Black Hole Entropy in String Theory"

%Douglas Kabat Pouliot Shenker, "D-Branes and Short Distances in  
String Theory"
\REF\dkps{M.~Douglas, D.~Kabat, P.~Pouliot and S.~Shenker,  
hep-th/9608024}

%W. Taylor, "D-brane Field Theory on Compact Spaces"
\REF\wt{W~.Taylor IV, hep-th/9611042}

%%%%%%%%%%%%%%%%%%%%%%%%%%%%%%%%%%%%%%%%%%%%%%%%%%%%%%%%%%%%%%%%%%%%%%%%%%%%%%

%
% *** Main body of paper ***
%
M-theory has provided a unifying principle to understand many of the
dualities of
string theory. For example, the strong-weak duality of type IIB  
theory is now
understood as a simple symmetry of permutation of two compactified
directions in the
11 dimensionlal framework. On the other hand, other symmetries such as
T-duality are
not easily undestood except in the limit of weakly coupled string theory.
 From the
M-theoretic viewpoint they are very mysterious. In this paper we  
will explore
T-duality from the vantage point of nonperturbative M-theory.

  Recently, the author together with Banks, Fischler and Shenker
 [\bfss] proposed a nonperturbative definition of uncompactified 11  
dimensional
M-theory as the large $N$ limit of  a supersymmetric  matrix quantum
mechanics. Let us
briefly review the construction.  We begin with type IIA string theory in
the limit of
weak coupling. The theory contains \dob \ in addition to other kinds of
objects. As
shown by Douglas, Kabat, Pouliot, and Shenker[\dkps], the \dob decouple
from everything
else as the string coupling tends to zero. The \dob in this limit  
are exactly
described by  the matrix quantum mechanics  which results from the  
dimensional
reduction of 10 dimensional SYM theory. If the number of \dob \ is $N$
(not to be
confused with the number of supersymmetries) then the  gauge symmetry is
$U(N)$.
Taking $N$ to infinity defines the infinite momentum frame description of
M-theory.

Our notations will be the same as in [\bfss]. The 9 transverse spatial
directions and time are labeled $t,Y^i   (i= 1...9)$  and the eleventh
direction is
$Y^{11}$. All lengths are measured in 11 dimensional planck units.  
The 11th
direction
is compactified on a circle of circumfrence $R$  which functions as  
an infrared
cutoff. It is eventually taken to infinity.

Let us consider the analogous construction to [\bfss] in the case  
where $K$
transverse
 directions are compactified on a K-torus.  It was argued in [\bfss] that
the matrix
model description of \dob \ must be replaced by a $K+1$ dimensional field
theory.  The
reason for this is the need to accomodate the virtual strings which  
connect
the \dob
\ and which are wound one or more times around the various cycles of the
torus. The
detailed construction  for the case $K=1$ was worked out  by Taylor  
[\wt].

 In this paper we will consider the case $K=3$.  The compact directions
 will be $Y^1,Y^2,Y^3=Y^a$.  The associated compactification scales  
will be
$L_a$.
$$
0<Y^a<L_a
$$
Let us begin by establishing the parameters of the 3+1 dimensional
 SYM theory. The base space for the theory will be labeled  by  
three spatial
coordinates $x^a$  which are compactified  according to
$$
0<x^a<S_a
$$
The coupling constant of the SYM theory will be $g$ and the Lagrangian is
$$
L=-{1 \over 4g^2}\int _0^S d^3x F_{\mu \nu}^2 + ....
\eqn\1
$$
To establish the values of the parameters $g$ and $S$ we
compare some
 energy scales of the SYM theory with energies that occur in
D0 brane physics. If we
work in the temporal gauge, and consider $x$ independent
 configurations the gauge
theory will have terms in the lagrangian of the form
$$
{S^1S^2S^3 \over 2g^2}(\dot A_a)^2
\eqn\2
$$
where the $A's$ are the homogeneous modes of the vector potential.
 The quanities $A_a S^a$ are related to Wilson loops and should be
treated as angular
variables. The conjugate variables are integer valued and the energy
corresponding to
\2 \ is a sum of terms of the form
$$
{(S^a)^2 g^2 n^2 \over 2 S^1S^2S^3}
\eqn\3
$$

Now compare this with the kinetic energy of a single D0 brane.
moving in the compact direction along the $a$ axis.
The D0 brane has longitudinal momentum $p_{11}$ given by $1/R$
 and transverse momentum $n/L^a$. The formula for its energy is  
$E={p_{a}^2
\over
2p_{11}}$ or
$$
E={n^2 R \over 2 (L^a)^2}
\eqn\4
$$
Equating eqs (3) and (4) gives
$$
{(S^a)^2 g^2  \over  S^1S^2S^3} = {R \over (L^a)^2}
\eqn\5
$$

Next consider the energy of a string wound once around  $L^a$.
 From the M-theoretic viewpoint, a string is a membrane wrapped  
around the 11th
direction and has tension equal to $R$ in 11 dimensional planck  
units. The
once-wound
string therefore has energy  ${L^a R \over{l_{11}^3}}$ where  
$l_{11}$ is the 11
dimensional planck length. This is to be equated  to the energy of a
``momentum mode"
in the SYM theory.  The energy of a single field quantum moving with a
single quanized
unit of momentum along the  $x^a$ direction is
${2\pi
\over S^a}$. Thus we find
$$
S^a = {2l_{11}^3 \pi \over {L^a R}}
\eqn\6
$$
Combining equations we find an expressin for $g$.
$$
g^2= {2 \pi {l_{11}^3} \over L^1 L^2 L^3}
\eqn\7
$$

Eqs 6 and 7 define the parameters of the SYM theory whose
large $N$ limit defines M-theory on a 3-torus.

Now let us change perspectives by thinking of the coordinate
$Y^3$ as the direction we would shrink to get  weakly coupled type IIA
theory.  Our
interest is in the constraints implied by the  T-duality which  
inverts the
size of the
2-torus $Y^1,Y^2$. This T-duality brings us to another IIA theory which
should be
equivalent to the original theory. Let us call the parameters of  
that theory
$\bar{L}^a $ and $\bar{l}_{11}$. Note that when T-duality is  
performed, the
11D planck
scale changes. This is because T-duality is defined by keeping fixed the
string scale
and 8-dimensional string coupling.  We find

$$\eqalign{
\bar{L}_1& ={l_{11}^3 \over L^3 L^1} \cr
\bar{L}_2& ={l_{11}^3 \over L^3 L^2}  \cr
\bar{L}_3& ={l_{11}^3 \over L^1 L^2}  \cr }
\eqn\8
$$
The new 11 dimensional planck length is given by
$$
\bar{l}_{11}^3={{l}_{11}^6 \over L^1L^2L^3 }
\eqn\9
$$
The coupling constant describing the new SYM theory
is given by the ``barred" analog of  eq (7).

$$
\bar{g}^2= {2 \pi {\bar{l}_{11}^3} \over \bar{L}^1 \bar{L}^2  \bar{L}^3}
\eqn\ten
$$

Using eqs (8) we find
$$
\bar{g}={2 \pi \over g}
\eqn\eleven
$$
One can also compute  compactification radii $\bar{S}$.  We find
$$\eqalign{
\bar{S}^1& =S^2 \cr
\bar{S}^2& =S^1  \cr
\bar{S}^3&=S^3  \cr }
\eqn\twelve
$$
Thus the new gauge theory is the same as the original except
 for two points. First, the two lengths $S^1$ and $S^2$ are interchanged.
That is a
trivial difference which clearly does not change the spectrum.  The
inversion of the
coupling in eq 10 is more serious. The point however is that the  
inversion is
the exact electric-magnetic duality symmetry of the $N=4$ (Here N
refers to
the number of supersymmetries) SYM theory. Evidently, the T-duality of
compactified
M-theory is insured by the S-duality of SYM.

Obviously there are a great many more relations between field  
theoretic and
string theoretic dualities that are implied by M(atrix) theory.

The Author would thank his collaborators, Tom Banks, Willy Fischler, and
Steve Shenker
for many valuable insights concerning every aspect of this paper. In
addition he is
grateful to Barak Kol, Arvind Rajaraman and Edi Halyo for very helpful
discussions.

\refout
\end